\documentstyle[12pt]{article}
\title{A REALISTIC INTERPRETATION FOR QUANTUM MECHANICS}
\author{Adonai S. Sant'Anna\thanks{e-mail: adonai@mat.ufpr.br}\\Dep. Matem\'atica - Universidade Federal do Paran\'a\\C.P. 19081, Curitiba, PR, 81531-990, Brazil}
\date{ }
\begin{document}
\maketitle

\noindent
We propose a realistic and nonlocal interpretation for quantum mechanics, which requires new mathematical, physical and philosophical foundations for space and time.

\section{Introduction}

	Before the famous experiment by Michelson and Morley in 1887, physicists believed that there should exist an ether in space, in order to explain the propagation of electromagnetic waves, by means of the mechanical theory of waves. However, that experiment showed that there is no relative motion of our planet with respect to a physical medium usually refered to as the ether.

	Now there is the belief that there should exist space-time as a medium which allows us to order physical events. The `new' ether is a collection of physical properties of a continuum space-time \cite{Einstein-91}. In this paper we propose that some phenomena in nature suggest that the continuum approach to space-time may be not the most appropriate. We work on the possibility that the world is, in some sense, atomistic. And space-time, as one of the constituents of the world, is also atomistic. We show that we may have causality without any time interva between two events.

	Some decades ago the japanese physicist T. Tati began a researche program about a description for physical theories where the concepts of space and time are not {\em primitive\/} notions \cite{Tati-64,Tati-83,Tati-86,Tati-87}. For a quick reference on his work see \cite{Tati-86}. For a complete description of the non-space-time picture of classical mechanics, electromagnetism, quantum mechanics and quantum electrodynamics see \cite{Tati-64}. Tati's objective is to solve a specific problem, namely, the divergences in quantum field theory. Tati argues that it is possible {\em to define\/} space and time in some physical theories, by using his theory. Space and time are fundamental concepts that should remain in classical physics. But in quantum field theories, the classical approach is meaningless. So, space and time do not exist at microscopic levels. Tati's formalism supports a theory of finite degree of freedom for quantum electrodynamics, which should eliminate the divergences that we mentioned above.

	We do not present in this paper all details of Tati's theory because we consider his formulation a bit confuse from the logico-mathematical standpoint. He does assume {\em causality\/} as a primitive concept, whatever does it really mean.

	So, our starting point is the intuition presented in Tati's work which, we believe, is somehow preserved in this paper (at least in principle).

	We venture to interpret Tati's work at our own risk, as it follows. Physical observations may be associated to elements of a discrete set which corresponds, intuitivelly speaking, to {\em state\/} measurement values. Each observation is related to another one by a {\em causal relation\/}. This causal relation may be expressed by equations which are very similar to numerical solutions of differential equations commonly used in physics. {\em Continuum} space-time, in theoretical physics, works just as a mathematical framework which allows us the use of standard differential and integral calculus. Space-time intervals, which are associated to elements of a {\em continuum\/}, should be regarded as `unknowables', if we use Tati's terminology. Such unknowables behave like hidden variables in the sense that we cannot actually measure {\em real\/} space-time intervals. Measurements do not have arbitrary precision. It should also be emphasized that even measurements of mass, position, momentum, etc., should be associated to elements of a discrete set of numbers, if we are interested to eliminate anthropomorphical notions like the {\em continuum} and real numbers.

	We recall the well known words by L. Kronecker: ``God made the integers, all the rest is the work of man.''

	In this paper we use a hidden variable formalism in order to present a realistic, deterministic and causal picture for quantum mechanics. We believe that there are at least two ways to get this goal: (i) a new mathematical formalism to the physical phenomena usually described by quantum theory and (ii) an adequate interpretation of the usual mathematics of quantum theory. We bet on the second one, since quantum mechanics has been very successful to explain with great accuracy several experimental facts.

	Recently, Ferrero and Santos \cite{Ferrero-97} claimed that ``physics has been constructed because three `philosophical' principles have been respected, namely, realism, locality, and consistency.'' By realism, the authors understand the ontological principle according to which ``there is an external world, independent of the subject and her/his mind that existed already before it and that will continue to exist when the subject perishes.'' This is the notion of realism that we adopt in this paper. By consistency Ferrero and Santos mean ``the necessity of no contradiction between the different theories of one concrete science, physics in this case.'' Again, we agree with them. Nevertheless, we should be more careful with respect to any locality principle in physics. Ferrero and Santos refer to Einstein's words \cite{Born-71}, according to whom physical objects are ``thoughtless arranged in a space-time continuum. An essential aspect of this arrangement of things in physics is that they may claim, at a certain time, to an existence independent of one another, provided that these objects `are situated in different parts of space'''.

	There are many experiments which support nonlocality, namely, self interference in the two-slit experiment \cite{Tonomura-89}, interaction-free measurements \cite{Kwiat-95} and quantum teleportation \cite{Bouwmeester-97}. But there is no experimental evidence which supports nonrealism in an objective manner. Chiao and Garrison \cite{Chiao-??} consider that locality contains an implicit form of realism, since space-time has physical properties independent of measurements performed by observers. So, we present in this paper a realistic picture for quantum mechanics, which violates Bell's inequalities. We pay a price for that: we abandon locality.

	Continuum space-time is just a mathematical tool, very useful to describe the dynamics of macroscopic phenomena. In other words, the continuum allows us the use of differential and integral calculus. Notwithstanding, the continuum gives rise to very serious conceptual and operational problems. It is well known, for example, that Turing machines do not provide an adequate notion for computability in the set of real numbers. If a discrete picture for physics is possible, some undecidable problems, which are very common in theoretical physics, should disappear.

	According to Dirac \cite{Dirac-58}: ``A measurement always causes the system to jump into an eigenstate of the dynamical variable that is being measured.'' We propose that quantum states are not linear combinations of eigenkets. Quantum states are eigenkets that are {\em oscillating\/}. Nontrivial linear combinations are physically meaningless, i. e., surplus structures.

\section{Non-Zero Spin Systems}

	In this Section we are first concerned with spin $\frac{1}{2}$ systems in the context of the Einstein-Podolsky-Rosen-Bohm (EPRB) experiment. Our notational features for EPRB are based on Sakurai's \cite{Sakurai-94}.

	We consider a two-electron system in a spin-singlet state. According to quantum mechanics (QM), if a given electron is known to be in an eigenket of spin in direction ${\bf a}$ with positive eigenvalue, the probability that the spin measurement of the other particle in the direction ${\bf b}$ yields a positive eigenvalue is

\begin{equation}
P({\bf a}+;{\bf b}+) = \sin^2\left(\frac{\theta_{ab}}{2}\right).\label{usual}
\end{equation}

	It is well known that equation (\ref{usual}) violates Bell's inequality.

	We propose to rewrite equation (\ref{usual}) as:

\begin{equation}
P({\bf a}+;{\bf b}+) = \sin^2\left(\pi n+\frac{\theta_{ab}}{2}\right),\label{unusual}
\end{equation}

\noindent
where $n = 0,1,2,3,...$. Such an equation is mathematically equivalent to the previous one (in the sense that it also violates Bell's inequality), and we interpret it as it follows.

\begin{description}

\item [First assumption] {\it Every quantum state of a spin  $\frac{1}{2}$ system is an {\em eigenket\/} of a spin  $\frac{1}{2}$ system.}

	This assumption is our first step towards a realistic picture for spin $\frac{1}{2}$ systems.

\item [Second assumption] {\it Every eigenket of a spin $\frac{1}{2}$ system is equivalent to an eigenket after rotation:

\begin{equation}
\vert\alpha\rangle = {\cal D}_{\bf a}(\phi)\vert\alpha_{before}\rangle,\label{phi}
\end{equation}

\noindent
where $\vert\alpha_{before}\rangle$ is the eigenket before the $\phi$-rotation about direction ${\bf a}$, and}

\begin{equation}
{\cal D}_{\bf a}(\phi) = \exp\left(\frac{-i S_{\bf a}\phi}{\hbar}\right),\label{Daphi}
\end{equation}

\noindent
{\it where $S_{\bf a}$ is the spin eigenvalue $\pm\frac{\hbar}{2}$ in the direction ${\bf a}$.}

\end{description}

\begin{description}

\item [Third assumption] {\it Spin eigenvalues oscillate between $+$ and $-$ sign, with frequency $\omega$. So, equation (\ref{phi}) should be rewritten as:

\begin{equation}
\vert\alpha\rangle = {\cal D}_{\bf a}(\omega t+\phi)\vert\alpha_{before}\rangle,\label{oscila}
\end{equation}

\noindent
where $t$ is time.}

\end{description}

	It is a logical consequence from these assumptions that if a given particle is known to be in an eigenket of spin in direction ${\bf a}$ with positive eigenvalue, the probability that the spin measurement of the other particle in the direction ${\bf b}$ yields a positive eigenvalue is

\begin{equation}
P({\bf a}+;{\bf b}+) = \sin^2\left(\frac{\omega}{2} t+\frac{\theta_{ab}}{2}\right),\label{omega}
\end{equation}

\noindent
where $\theta_{ab}$ is the angle between directions ${\bf a}$ and ${\bf b}$. 

\begin{description}

\item [Fourth assumption] {\it There is a real time unit $\tau>0$ where $\omega = 2\pi$.}

\item [Fifth assumption] {\it There is no time interval $t>0$ such that $t<\tau$.}\footnote{Since $\tau$ is a real number, we make no reference to infinitesimals or non-standard analysis.}

\end{description}

	If we adopt $\tau$ as our time unit, equation (\ref{unusual}) can be easily derived from equation (\ref{omega}), where $n$ stands for the discrete instants of time granted by the fifth assumption.

\begin{description}

\item [Sixth assumption] {\it To perform a measurement of a spin $\frac{1}{2}$ system means to determine the eigenket $\vert\alpha\rangle$ at an instant $t$ of time, according to the dynamics described by equation (\ref{oscila}).}

\end{description}

	This last assumption deserves further explanation. If the phase angle $\phi$ is zero, equation (\ref{oscila}) says that the ket $\vert\alpha\rangle$ oscillates between $+$ and $-$, which apparently contradicts QM. Nevertheless, note that equation (\ref{omega}) says that if $\phi$ is zero, then $P({\bf a}+;{\bf b}+) = 0$. Hence, we conclude by that that the choice of $\phi$ determines our chances of getting a specific eigenvalue for spin.

	It is important to say that the sixth assumption does {\em not\/} say that measurements cease the oscillation postulated by equation (\ref{oscila}). Measurements just define the value for the phase $\phi$.

	From equation (\ref{Daphi}) it is easy to prove that for spin $s$ systems we have:

\begin{equation}
P({\bf a}+;{\bf b}+) = \sin^2\left(2\pi s\, n + s\theta_{ab}\right).\label{geral}
\end{equation}

	Hence, if $s$ is either an integer or a $\frac{1}{2}$ multiple of an integer, it is easy to prove that equation (\ref{geral}) violates Bell's inequality. In the case of spin zero particles, rotations are not observable.

\section{Hidden Variables and the Cat Paradox}

	According to QM, after a spin measurement in the $z$ direction in a $xyz$ coordinate system, the ket state of, e.g., one electron may be written as:

\begin{equation}
\vert S_x\rangle = \frac{1}{\sqrt{2}}(\vert +\rangle + \vert -\rangle),\label{linear}
\end{equation}

\noindent
where $\vert S_x\rangle$ stands for the ket state of the spin in the $x$ direction.

	In our picture we consider that spin eigenvalues are oscillating between $+$ and $-$ sign, with a certain frequency $\omega$, as it can be seen in the second assumption of the previous Section. So, linear combinations of eigenkets are useless in our formulation. The lack of realism in linear combinations is no longer a problem in our approach. {\em The quantum physical state of a system is an oscillating eigenket.}

	Suppose, now, we put a living cat in a cage with a spin $\frac{1}{2}$ particle (say, an electron), a spin measurement device, and a gun pointed to the heart of the cat.  If the spin measurement yields a positive eigenvalue in a given direction, the gun is activated, and it will kill the cat.  On the other hand, if a spin measurement yields a negative value, nothing happens and the cat will be still alive.  The question is: What is the state of the cat after, say, 10 minutes, if no spin measurement is performed?\footnote{This {\em gedanken\/} experiment is a variant of the well known cat paradox first stated by Erwin Schr\"odinger in 1935.}

	According to the usual interpretation of QM, the state of the cat is a linear combination of {\em dead} and {\em alive}. According to our picture the cat is always dead, after some time interval, since death is an irreversible process. But the gun pointed to the animal is discretely shooting.

	Nevertheless, we do not know the value of $\omega$ in usual units of time. It shall be, we conjecture, a very large number. So, equation (\ref{linear}) just translates our ignorance on the value of the hidden variable $\omega$. In this sense, $\tau$ is also a hidden variable.

\section{Position and Momentum}

	By analogy to the previous case of non-zero spin systems, we consider as our first assumption the following:

\begin{description}

\item [PM-1] {\it Every quantum state of position is an eigenket of position, and every eigenket of position is equivalent to an eigenket after translation:

\begin{equation}
\vert x\rangle = {\cal T}(\delta)\vert x_{before}\rangle,\label{pm}
\end{equation}

\noindent
where $\vert x_{before}\rangle$ is the eigenket before the translation by $\delta$, and ${\cal T}(\delta)$ is the translation operator.}

\item [PM-2] {\it Position eigenvalues oscillate according to the following equation,

\begin{equation}
\vert x\rangle = {\cal T}(\delta_n)\vert x_{before}\rangle,
\end{equation}

\noindent
where $n$ is a natural number which corresponds to the same time unit in the previous Section.}

\item [PM-3] $\delta_n$ {\it is a sequence whose set of images $Im (\delta_n)$ is given as follows:

\begin{equation}
Im (\delta_n)\subseteq \left[ x'-\frac{\Delta}{2},x'+\frac{\Delta}{2}\right],
\end{equation}

\noindent
i.e., $Im (\delta_n)$ is a subset of the real interval with center at $x'$ and radius $\Delta/2$.}

\item [PM-4] {\it To perform a measurement of position means to change the value of $\Delta$.}

\end{description}

	What is the essential difference between the sixth assumption for spin $\frac{1}{2}$ particle systems and this last axiom for position measurements? Spin measurements yield just one between {\em two\/} possible values. But, position measurements yield, in principle, to one among infinite possibilities. So, we are seriously constrained to technological difficulties to get just {\em one\/} value for position measurements, since $\delta_n$ shall be very small. Axiom {\bf PM-4} takes into account these technical difficulties.

	Since we assume that $\delta_n$ is very small, we can assume also that

\begin{equation}
{\cal T}(\delta_n) = 1 - \frac{i}{\hbar}{\bf p}\delta_n,
\end{equation}

\noindent
where $i = \sqrt{-1}$, $\hbar$ is the reduced Planck's constant, ${\bf p}$ is the momentum operator and $1$ is the unitary operator.

\section{Time Evolution}

	It is quite obvious that in our picture the time evolution operator ${\cal U}(t,t_0)$ does have a discrete spectrum for eigenvectors and eigenvalues. That does not invalidate Schr\"odinger equation, since the time unit $\tau$ is assumed very small. So, the continuous works just as an idealization or an approximation. Such approximations are very common in quantum physics. See, for example, the Casimir effect \cite{Milonni-94}.

	It is well known that a monochromatic electromagnetic field is mathematically equivalent to a harmonic oscillator of the same frequency, whose Hamiltonian allows us only a discrete distribution of energy levels. Since, loosely speaking, the Hamiltonian is the generator of time evolution and energy is quantized, why cannot we consider time as quantized?

	We recognize that this is not a mathematically rigorous argument. But we consider this heuristics as some kind of motivation for time quantization in quantum mechanics.

	One natural question concerns a relativistic equation for the electron, under similar assumptions. We hope to present our version for Dirac's equation in future papers as soon as possible.

\section{Final Remarks}

	Actions-at-a-distance do not occur only in quantum theory. Newtonian forces are also instantaneous. In contrast, Einstein's General Theory of Relativity does not allow actions-at-a-distance since it takes into account the existence of a gravitational field, which implies the existence of gravitational waves. Nevertheless, such waves were never detected, until now. There is a natural resistence against the possibility that space-time is just an illusion, a mathematical trick which allows us the use of ordinary calculus in theoretical physics.

	Another interesting point referes to the discrete time proposed in the fourth assumption for spin $\frac{1}{2}$ systems. In the beginning of the history of quantum theory, some authors refered to the famous Stern-Gerlach experiment as a phenomenon of {\em space quantization\/}. Here, we suggest a space-time quantization.

\section{Acknowledgments}

	We acknowledge with thanks some useful criticisms by Osvaldo Pessoa Jr in an earlier version of this paper.

\end{document}